\documentclass[aps,prl,reprint,superscriptaddress,twocolumn]{revtex4}
\usepackage[dvips]{graphicx}
\usepackage{wrapfig}
\usepackage{subfigure}
\usepackage{bm}
\usepackage{color}
\usepackage{here}
\usepackage{amsfonts} 
\usepackage{latexsym}

\begin{document}

\title{
Stability of Spinmotive Force in Perpendicularly Magnetized Nanowires under High Magnetic Fields
}

\author{Y. Yamane}%$^{\ast\ 1,2}$
\affiliation{Advanced Science Research Center, Japan Atomic Energy Agency, Tokai 319-1195, Japan}
\affiliation{Institute for Materials Research, Tohoku University, Sendai 980-8577, Japan}
\author{J. Ieda}%$^{1,3}$,
\affiliation{Advanced Science Research Center, Japan Atomic Energy Agency, Tokai 319-1195, Japan}
\affiliation{CREST, Japan Science and Technology Agency, Tokyo 102-0075, Japan}
\author{S. Maekawa}%$^{1,3}$}
\affiliation{Advanced Science Research Center, Japan Atomic Energy Agency, Tokai 319-1195, Japan}
\affiliation{CREST, Japan Science and Technology Agency, Tokyo 102-0075, Japan}

%===================================================================================================
%           Abstract
%===================================================================================================

\begin{abstract}
Spinmotive force induced by domain wall motion in perpendicularly magnetized nanowires is numerically demonstrated.
We show that using nanowires with large magnetic anisotropy can lead to a high stability of spinmotive force under strong magnetic fields.
We observe spinmotive force in the order of tens of $\mu$V in a multilayered Co/Ni nanowire and in the order of several hundred $\mu$V in a $L1_0$-ordered FePt nanowire;
the latter is two orders of magnitude greater than that in permalloy nanowires reported previously.
The narrow structure and low mobility of a domain wall under magnetic fields in perpendicularly magnetized nanowires permits downsizing of spinmotive force devices.
\end{abstract}

\maketitle

%===================================================================================================
%           Introduction
%===================================================================================================

The mutual interaction between spin current and magnetization is a key phenomenon in the field of spintronics\cite{maekawa}.
Spin current induces magnetization dynamics, i.e., spin-transfer-torque\cite{stt,stt2,stt3,stt4}, which has been extensively studied through current-induced domain wall (DW) motion, primarily in permalloy (NiFe) nanowires with in-plane magnetization\cite{nife1,nife2,nife3,nife4,nife5}.
Recently, it has been reported experimentally\cite{pma1,pma2} and theoretically\cite{pma3,pma4} that nanowires with perpendicular magnetic anisotropy (PMA) are more favorable for spin-transfer-torque devices because they exhibit DW motion at a lower threshold current.
On the other hand, the same interaction mediates energy-transfer from magnetization into conduction electrons, i.e., spinmotive force (SMF)\cite{barnes}.
Like the spin-transfer-torque\cite{parkin} and other spin-related electromotive forces\cite{zutic1,zutic2}, SMF offers a promising functionality for future spintronic devices;
%As well as the spin-transfer-torque\cite{parkin} or other electromotive forces of spin origin\cite{zutic1,zutic2}, SMF is a promissing phenomenon for future spintropic devices;
e.g., it can be used to read out magnetic information\cite{barnes2}.
SMF has been primarily demonstrated in NiFe\cite{yang,ohe,shape,duine2,comb,hayashi}, and only few reports have demonstrated SMF in PMA materials\cite{martin}.
However, the soft magnetic properties of NiFe could occasionally be a major disadvantage for an SMF demonstration;
certain magnetization structures are easily disturbed by a strong magnetic field, leading to an increased instability in SMF signals.
Therefore, SMF induced by DW motion in NiFe nanowires is limited to a few $\mu$V\cite{yang,hayashi}.

In this paper, the SMF induced by DW motion in PMA nanowires is numerically investigated.
The above-mentioned problem in the NiFe nanowires is resolved by using PMA nanowires, which show rigid DW motion because of their high magnetic anisotropy.
It has been shown that materials with the high PMA are effective in achieving a large SMF.
We numerically observe SMF in the order of tens of $\mu$V in a multilayered Co/Ni nanowire and in the order of several hundred $\mu$V in a $L1_0$-ordered FePt nanowire with the very large PMA.
In addition, the narrow structure and low mobility of a DW in PMA nanowires are effective in downsizing of devices using SMF.

%===================================================================================================
%           Analytical discussion
%===================================================================================================

Let us begin by assessing SMF induced by DW motion in NiFe nanowires.
It has been pointed out that SMF is generally produced by the spin electric field\cite{volovik,duine,tser,yamane,shibata};
\begin{equation}
\bm{E}_\mathrm{s}=-\frac{P\hbar}{2e}\bm{m}\cdot\left(\frac{\partial\bm{m}}{\partial t}\times\nabla\bm{m}\right)
\label{e}\end{equation}
where $\bm{m}$ is the unit vector parallel to the direction of local magnetization of the ferromagnet, $P$ is the average spin polarization of the conduction electrons, and $e$ is the elementary charge.
It is required that the magnetization depend on both time and space.
By assuming a one-dimensional model for the DW structure and by using Eq.~(1), we find that the SMF, $V$, associated with field-induced DW motion is proportional to the applied magnetic field $\mu_0H$\cite{yang,hayashi,duine2,shape};
\begin{equation}
V=\frac{P\hbar\omega}{e}
\label{v}\end{equation}
where $\omega=\gamma \mu_0 H$ is the Larmor angular frequency, in which $\gamma$ is the gyromagnetic ratio and $\mu_0$ is the magnetic constant.
Yang \textit{et al.}\cite{yang} and Hayashi \textit{et al.}\cite{hayashi} experimentally demonstrated DW-motion-induced SMF in NiFe nanowires.
NiFe has been widely used to study magnetization dynamics because of its extremely soft magnetic properties.
However, such soft magnetic properties restrict the applicability of Eq.~(\ref{v}), which implicitly assumes a single DW.
When the applied field reaches $\sim 10$ mT, the structure of the DW in the NiFe nanowires begins to change from moment to moment, accompanying multiple vortex and anti-vortex core nucleation, and the DW behaves quite nonlinearly with respect to the applied field\cite{hayashi2}.
Such magnetization dynamics causes multiple spikes in the real-time observed SMF signal, while the time-averaged voltage is still well described by Eq.~(\ref{v})\cite{hayashi}.
As the applied field is further increased, another DW is created at the end of the NiFe wire and it propagates in the direction oppsite to that of the original DW.
The SMF is cancelled out becasue of this counterpropagating DW motion.
Therefore, for SMF generation, the magnetic field applied to NiFe nanowires is limited to tens of mT.
Thus far, experimentally achieved SMF has been limited to not more than a few $\mu$V\cite{yang,hayashi}.

The above-mentioned limitation is resolved by using PMA nanowires, which have a very rigid DW structure.
A spatially well-confined single DW structure is maintained until the applied magnetic field exceeds the crystalline anisotropic magnetic field.
The critical magnetic field $\mu_0H_\mathrm{c}$ is roughly estimated as $2K_\mathrm{u}/M_S$, where $K_\mathrm{u}$ and $M_S$ are the uniaxial magnetic anisotropy and the saturation magnetization, respectively.
In PMA nanowires, the demonstration of SMF generated by the field-induced DW motion is possible up to $\mu_0H_\mathrm{c}$.
Therefore, a larger SMF is expected in materials with higher PMA.

%===================================================================================================
%           Figure 1 (a)
%===================================================================================================

\begin{figure}[t]
    \centerline{\includegraphics[width=80mm]{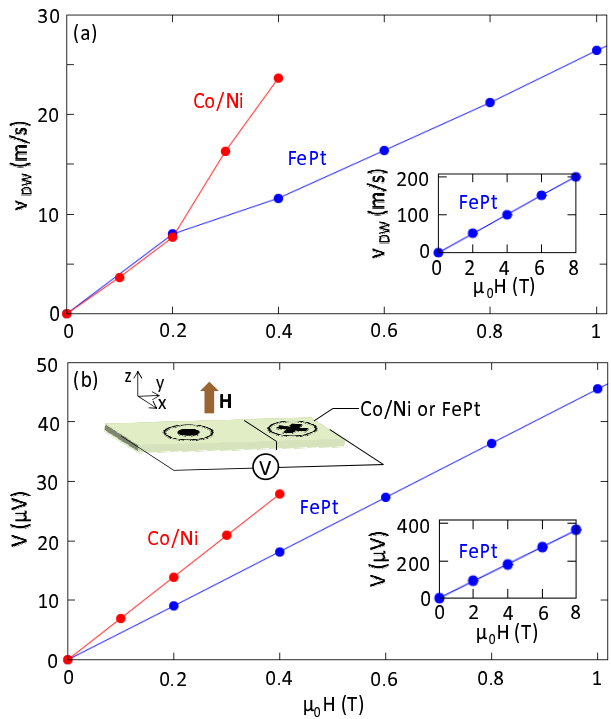}}
    \caption{  Time averaged (a) DW velocity and (b) SMF between the wire edges for Co/Ni and FePt nanowires as a function of the magnetic field applied along the out-of-plane direction.
               A high-field regime for the FePt nanowire is shown in the bottom-right insets.
               The top-left inset in (b) shows a schematic of the present system.
            }
    \label{fig3}
\end{figure}

The field-induced DW motion in PMA nanowires, multilayered Co/Ni, and $L1_0$-ordered FePt, is investigated by numerically solving the Landau-Lifshiz-Gilbert equation
\begin{equation}
\frac{\partial\bm{m}}{\partial t}  =   -  \gamma \bm{m} \times \bm{H}_\mathrm{eff} + \alpha \bm{m} \times \frac{\partial\bm{m}}{\partial t},
\label{llg}\end{equation}
using the public OOMMF code\cite{oommf}.
Here $\alpha$ is the Gilbert damping constant, and $\textrm{\boldmath $H$}_\mathrm{eff}$ is the effective magnetic field including the external, exchange, demagnetizing, and crystalline anisotropic fields.
Material parameters are as follows\cite{param1,param2,param3}:
$M_S=6.8\times 10^5$ A/m, $K_\mathrm{u}=4\times 10^5$ J/m$^3$, $\alpha=0.02$, and $P=0.6$ for the Co/Ni nanowire;
$M_S=1.035\times 10^6$ A/m, $K_\mathrm{u}=5\times 10^6$ J/m$^3$, $\alpha=0.1$, and $P=0.4$ for the FePt nanowire;
and $\gamma=1.76\times 10^{11}$ Hz/T and the exchange stiffness $A=10^{-11}$ J/m for both nanowires.
The width, length, and thickness of each nanowire are $40$ nm, $1$ $\mu$m, and $4$ nm, respectively.
We choose unit cell sizes of $1\times 1\times 4$ nm$^3$ for the Co/Ni nanowire and $0.4\times 0.4\times 4$ nm$^3$ for the FePt nanowire.
Initially, a Bloch-type DW is prepared in the middle of each nanowire.
The width of the DW in Co/Ni and FePt is approximately $10$ nm and $2$ nm, respectively.

Figure 1 (a) shows the DW velocity $\mathrm{v}_{DW}$\cite{vdw} in each wire as a function of the magnetic field applied along the out-of-plane direction.
A high-field regime for FePt is shown in the bottom-right inset.
Numerically observed $\mu_0H_\mathrm{c}$ for Co/Ni and FePt is $0.4$ and $8$ T, respectively, which is consistent with the estimated values $2K_\mathrm{u}/M_S = 1.17$ and $9.66$ T.
Above $\mu_0H_\mathrm{c}$ multiple domains appear and a single DW velocity can no longer be defined.
This multiple-domain regime will be discussed in more detail elsewhere.
In the Co/Ni nanowire, the DW velocity becomes nonlinear above $0.2$ T, whereas in the FePt nanowire with a larger PMA, the DW motion stays rigid and linear up to $8$ T.
The DW mobilities in each wire agree well with the one-dimensional model\cite{slonc}, $\mathrm{v}_{DW}/H=(\alpha+\alpha^{-1})^{-1}\gamma\sqrt{A/K_\mathrm{u}}\simeq 18$, and $25$ ms$^{-1}$T$^{-1}$ for Co/Ni and FePt, respectively.
These values are much lower than that of NiFe nanowires, which shows DW velocity of several hundred m/s with $\mu_0H\sim 0.01$ T.
This implies that PMA nanowires of a few micrometers in length are sufficient to demonstrate SMF induced by DW motion, thus enabling downsizing of SMF devices.
Note that the magnitude of SMF is independent of the DW velocity, see Eq.~(\ref{v}).

%===================================================================================================
%           Figure 1 (b)
%===================================================================================================

\begin{figure}[t]
    \centerline{\includegraphics[width=70mm]{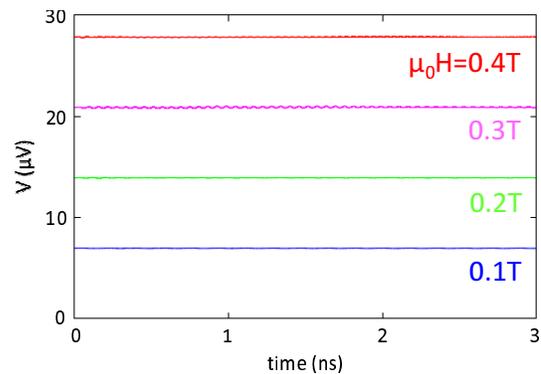}}
    \caption{  Time evolution of SMF between the ends of the Co/Ni nanowire under several magnetic fields.
               These signals are almost constant in time.
               The angular frequencies of the small oscillations of amplitude $\sim 0.1$ $\mu$V in each signal are twice the Larmor angular frequency, $2\gamma\mu_0 H$.
            }
    \label{fig3}
\end{figure}

A measurable voltage between two given points is defined as the spatial difference of the spin-independent electric potential $\phi(\bf{x})$ between the points, calculated from the Poisson equation $\nabla^2\phi=-\nabla\cdot\bf{E}_\mathrm{s}$\cite{yang,ohe,duine2,yamane}.
Figure 1 (b) shows the time-averaged SMF, $V$, between the ends of the wire as a function of the magnetic field for each wire.
The top-left and bottom-right insets show a schematic of the present system and the voltage for the FePt nanowire in a high-field regime, respectively.
These numerical results well reproduce the analytical prediction obtained using Eq.~(\ref{v});
the voltage is proportional to the applied magnetic field $\mu_0H$, its slope is determined by the spin polarization $P$, and the DW velocity is irrelevant to its magnitude.
A material with higher PMA realizes larger SMF because of the larger critical field $\mu_0H_\mathrm{c}$.
This is our main claim.

%===================================================================================================
%           Figure 2
%===================================================================================================

Next, we examine the time and spatial dependence of SMF.
Hereafter, we will focus on the Co/Ni nanowire as the result is qualitatively similar for the FePt nanowire.
In Fig.~2, we show the time evolution of SMF in the Co/Ni nanowire under several magnetic fields.
The signals are stable, i.e., almost constant in time, because of the rigid DW motion.
This result is in contrast with those of the NiFe nanowire, in which multiple spikes appeared in the observed SMF signal\cite{hayashi}.
We remark that the small oscillations of amplitude $\sim 0.1$ $\mu$V in the SMF signals, shown in Fig.~2, reflect the coherent precession of the DW\cite{coher};
the SMF is proportional to the angular frequency of the magnetization around the magnetic field, see Eq.~(\ref{v}), which now depends on $m_x$ and $m_y$.
The angular frequency of the coherent precession is twice the Larmor angular frequency, $2\omega=2\gamma\mu_0 H$.

%===================================================================================================
%           Figure 3
%===================================================================================================

Figure 3 (a) shows the magnetization configuration near the DW in the Co/Ni nanowire at $t=1.5$ ns and $\mu_0 H=0.4$ T.
The color code represents the out-of-plane component of the magnetization.
The DW is confined to within a range of approximately $10$ nm, even under the motion in such a relatively large magnetic field.
Figure 3 (b) shows the corresponding electric-potential profile.
The spin electric field (\ref{e}), which reflects the time and spatial dependent magnetization dynamics, appears only in the vicinity of the moving DW.
It can be obsserved that the potential drop actually occurs in the narrow confine.
\begin{figure}[t]
    \centerline{\includegraphics[width=70mm]{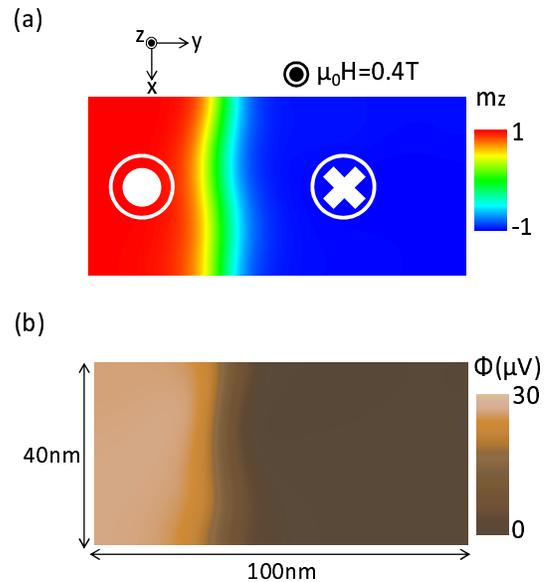}}
    \caption{  (a) The magnetization configuration around the DW in the Co/Ni nanowire at $t=1.5$ ns with $\mu_0 H=0.4$ T and (b) the corresponding electric-potential distribution.
            }
    \label{fig3}
\end{figure}

%===================================================================================================
%           Summary
%===================================================================================================

In summary, we have investigated the SMF caused by field-induced DW motion in PMA nanowires.
By using PMA materials, the problem in NiFe nanowires was resolved;
a high stability of SMF was achieved under strong magnetic fields.
Materials with a high PMA are favorable for achieving a large SMF.
We observed SMF in the order of tens of $\mu$V in a multilayered Co/Ni nanowire and in the order of several hundred $\mu$V in a $L1_0$-ordered FePt nanowire.

%===================================================================================================
%           Acknowledgments
%===================================================================================================

The authors are grateful to T.~Seki in Institute for Materials Research, Tohoku University for valuable comments on this work.
This research was supported by a Grant-in-Aid for Scientific Research from MEXT, Japan.

%===================================================================================================
%           References
%===================================================================================================


\begin{thebibliography}{99}
\bibitem{vdw}      The DW velocity is defined as $d/t$ m/s, where $d$ is the distance that the DW center travels over the time $t=3$ ns.
                   Since the DW is now narrow enough, the DW center is clearly defined.
\end{thebibliography}
\end{document}